\documentstyle[prb,preprint,aps]{revtex}
\begin{document}
\draft

\title{Phase separation and the existence of superconductivity 
in a one-dimensional copper-oxygen model}

\author{A. W. Sandvik}
\address{National High Magnetic Field Laboratory, Florida State
University, 1800 E. Paul Dirac Dr., Tallahassee, Florida 32306 }

\author{A. Sudb\o}
\address{Department of Physics, Norwegian University of Science and 
Technology, N-7034 Trondheim, Norway}

\date{February 21, 1996}

\maketitle

\begin{abstract}
The phase separation instability occurring with increasing nearest-neighbor 
repulsion $V$ in a two-band Hubbard model (CuO chain) is 
discussed. Quantum Monte Carlo simulations indicate that this transition is 
associated with a level-crossing if the filling fraction $\langle n\rangle$ 
is close to $1$ (half-filled lower band). Spin-density-wave fluctuations
then dominate before phase separation. Superconducting fluctuations 
dominate only at considerably higher 
doping levels.
\end{abstract}
\vskip5mm
\vfill\eject

It has been suggested that superconductivity in strongly correlated electron
systems is favored close to regions in parameter space where phase
separation occurs.\cite{emery1,dagotto1} This hypothesis is based on 
the notion that the 
fluctuations that cause the system to separate into high- and 
low-density regimes can lead to pairing if they are not quite strong
enough to lead to phase separation. An explicit example of this mechanism
is observed in the one-dimensional (1D) $t$-$J$ model.\cite{ogata} 
For small ratios $J/t$ this
system is a Luttinger liquid,\cite{haldane,schulz} 
characterized by a single exponent
$K_\rho$ which governs the decay of the spin and charge correlations,
as well as the singlet and triplet pair correlations.
A system with $K_\rho > 1$ has dominant singlet pairing fluctuations,
whereas spin-density-wave fluctuations dominate if $K_\rho < 1$.
The exponent is related to the compressibility $\kappa$ and the 
Drude weight $D$ according to\cite{schulz}
\begin{equation}
K_\rho^2 = \pi \kappa D /4 .
\label{krho}
\end{equation}
For the 1D $t$-$J$ model, the compressibility increases 
smoothly with increasing $J/t$ and diverges as the system
phase separates for $J/t \approx 3$.\cite{ogata} The Drude weight remains 
finite across the phase separation boundary, vanishing only for slightly
larger $J/t$ (the high-density phase is initially conducting).\cite{hellberg}
$K_\rho$ hence diverges 
at the phase separation boundary, and there is always 
a regime with dominant pair correlations prior to phase 
separation. Emery, Kivelson and Lin have proposed that the same fluctuations
that lead to phase separation may induce pairing 
also in the 2D $t$-$J$ model.\cite{emery1} 
Dagotto {\it et al.} have recently conjectured that the existence 
of superconductivity in the neighborhood of phase separation is a 
generic feature of strongly correlated electron systems.\cite{dagotto1}

Here we study a 1D two-band version \cite{sudbo} of the 2D 
three-band Hubbard model suggested as a model of the CuO$_2$ planes
of the high-T$_{\rm c}$ cuprate superconductors.\cite{emery2,varma1}
We fix the single-particle potentials and the on-site Cu 
repulsion at values appropriate for the high-T$_{\rm c}$
cuprates, and study the behavior versus the near-neighbor
repulsion $V$. We consider fillings $\langle n\rangle$ 
ranging from $1$ (half-filling) to $2$ (completely filled lower band).
Using a quantum Monte Carlo (QMC) technique, we find that phase 
separation close to half-filling is a result of a level crossing: A band 
of already phase separated states cross the Luttinger liquid ground state
as $V$ is increased. The system has dominant spin-density-wave fluctuations
before phase separation. On the other hand, for $\langle n\rangle$
close to $2$, fluctuations in the Luttinger liquid state do lead
to dominant pair correlations, and also appear to cause the
phase separation. Hence, we have an example of a system where
the interactions causing phase separation do or do not
also induce superconductivity, depending on the filling.

The Cu-O chain model is defined by the Hamiltonian
\begin{eqnarray}
\hat H = & & 
 -t\sum\limits _{\sigma} \sum\limits _{i=1}^{2L}
\bigl [c^+_{i+1,\sigma}c_{i,\sigma} + c^+_{i,\sigma}c_{i+1,\sigma}\bigr ] 
  \nonumber \\
& & + \sum\limits _{i=1}^{2L} \bigl [
\Delta_i n_i + U_i n_{i,\uparrow} n_{i,\downarrow} + V n_{i}n_{i+1} \bigr ].
\label{model}
\end{eqnarray}
Here $c^{+}_{i,\sigma}$ is a creation operator for a spin-$\sigma$
particle on site $i$, with odd and even $i$ corresponding to copper
$d$ and oxygen $p$ orbitals, respectively, on a lattice with $L$ unit
cells ($N=2L$ sites), and $n_i = n_{i,\uparrow} + n_{i,\downarrow}$. 
$\Delta_i$ and $U_i$ denote the single-particle 
potential and Hubbard repulsion, with $\Delta_i = 0, U_i=U_d$ for odd 
$i$ and $\Delta_i = \Delta, U_i=U_p$ for even $i$. 
We work in the hole representation, i.e. increasing the filling above
$\langle n\rangle =1$ corresponds to hole doping the insulating 
cuprate parent compounds. Typical estimates for the parameters
appropriate for a CuO$_2$ plane are, in units of $t$,
 $\Delta=2.5-3$, $U_d=6-10$, $U_p=2-4$.\cite{parameters} $V$ is 
difficult to estimate. For the 1D Cu-O model studied here we choose 
$U_d=6$, $U_p=0$, and $\Delta=2$, which gives a robust 
charge-transfer insulator at half-filling, and regard $V$ as a 
variable.

We use a QMC method based on stochastic series
expansion \cite{qmc} 
(a generalization of Handscomb's technique \cite{handscomb}). 
This method is free from errors associated with the Trotter
decomposition used in standard methods.\cite{worldline} 
We have calculated the spin and charge structure factors
\begin{equation}
S_{\rho,\sigma}(q)=\frac{1}{L}~\sum_{k,j} e^{iq(k-j)}
\langle N_{\rho,\sigma}(k)N_{\rho,\sigma}(j)\rangle ,
\end{equation}
and the corresponding static susceptibilities $\chi_{\rho,\sigma}(q)$.
Here, $N_{\rho,\sigma}(j)$ is the charge ($\rho$) or spin ($\sigma$)
in unit cell $j$. We have also calculated current-current correlation 
functions, from which the Drude weight can be obtained.\cite{stiffness} 
Using these quantities we have extracted the $V$-dependence of the 
low-energy parameters, $K_\rho$ in particular, at several fillings. 
Details of the computations are discussed in Ref.~\onlinecite{cuolong}. 
Here we focus on the mechanism causing phase separation in this model, 
and the issue of whether the system possesses dominant pairing fluctuations 
in the neighborhood of phase separation.  

For fillings $1 < \langle n\rangle < 2$ a
large $V$ causes the system to separate into high-density (HD) and 
low-density (LD) phases. This can be easily seen in the limit $V \to \infty$,
where nearest-neighbor sites cannot be simultaneously occupied. At 
half-filling all $d$ sites are then singly occupied, and the $p$ sites are 
all empty. Adding particles to this state, the $V$ repulsion is minimized 
by the formation of a HD regime with doubly occupied $p$ sites and empty $d$ 
sites, and an LD regime which remains as the half-filled system with zero 
$p$ occupation. Upon lowering $V$, the particles acquire some mobility, 
but the HD and LD phases initially remain at insulating densities 
$\langle n\rangle_{HD} =2$ and $\langle n\rangle_{LD} =1 $, respectively. 
For even lower $V$, our 
results discussed in Ref.~\onlinecite{cuolong} show that particles from the 
phase boundary first evaporate into the LD phase. The system is
still phase separated for some range of $V > V_{PS}$, but the 
LD phase is conducting, with $\langle n\rangle _{LD} > 1$. 

To shed further light on the details of the phase-separation instability,
and in what $(V,\langle n\rangle)$ regime it occurs,
it is useful to study local quantities that can be expected to change 
rapidly as $V$ approaches $V_{PS}$, such as the kinetic energy, the 
occupation of $p$ sites, and the number of doubly occupied $p$ sites. 
These are shown versus $V$ for a filling $\langle n\rangle = 1.25$
in Fig.~\ref{local}. The simulations were carried out at temperatures
low enough to give essentially ground state results (inverse
temperatures $\beta = t/T$ up to $128$ were used). For $V \alt 2.5$ 
there is little dependence on $L$. We have 
verified that the system is a Luttinger liquid in this regime, by checking 
the consistency of relations among the low-energy parameters,\cite{schulz} 
such as Eq.~(\ref{krho}). As $V$ is increased, there is a size dependent
range where there is a rapid change in all the quantities shown. It is
natural to associate this with phase separation. Indeed, the behavior
of the charge structure factor also changes qualitatively in the
same range. For example for $L=32$, $S_\rho (q)$ at $V=2.6$ decreases
to zero as $q \to 0$, whereas it is strongly peaked at $q=q_1=2\pi/L$ 
for $V=2.7$. The latter behavior is a clear sign of phase separation. The size 
dependence of $V_{PS}$ is probably related to the existence of 
finite boundaries between the LD and HD phases, where the $V$ interactions
are not avoided as efficiently as in the bulk of the phases, and which 
are relatively large in small systems. 

A striking feature is that all quantities appear to change discontinuously
at $V_{PS}$ for the larger systems. We propose that this 
behavior is associated with a level spectrum schematically outlined in
Fig.~\ref{levels}: In addition to the Luttinger liquid and its standard 
spin and charge excitations, there is a band of states corresponding to 
a phase separated system. These states are present even for $V$ significantly
lower than $V_{PS}$, but are then high in energy. As $V$ approaches 
$V_{PS}$ they approach, and cross, the Luttinger liquid ground state.

Assume that there are indeed two sets of low-energy 
states $\lbrace LL \rbrace$ and $\lbrace PS \rbrace$ 
corresponding to a Luttinger liquid and a phase separated system, 
respectively, as illustrated in Fig.~\ref{levels}. For a large system, states 
belonging to different sets will in general be very different, in the 
sense that their expansions in terms of eigenstates of the 
$r$-space number operators are dominated by different states, i.e.~if 
a state belonging to $\lbrace LL \rbrace$ has a large expansion 
coefficient $a_i$ for a state $\psi_i$, all states  belonging to 
$\lbrace PS \rbrace$ have small coefficients $a_i$. This can be expected
to have consequences for a simulation carried out in the $r$-space basis. 
Typical low-temperature configurations can then be
uniquely associated either with $\lbrace LL \rbrace$ or $\lbrace PS \rbrace$.
A configuration belonging to $\lbrace LL \rbrace$ will require a
large modification in order to ``tunnel'' into $\lbrace PS
\rbrace$, and vice versa. Since the simulation proceeds 
in steps of small modifications of the configurations, one can expect 
tunneling between the sets to be rare. Consider the ``partial partition 
functions''
\begin{equation}
Z_{LL} = \sum\limits_{i \in \lbrace LL\rbrace} {\rm e}^{-\beta E_i},\qquad
Z_{PS} = \sum\limits_{i \in \lbrace PS\rbrace} {\rm e}^{-\beta E_i},
\nonumber
\end{equation}
and the corresponding free energies $F_{LL} = -\ln{(Z_{LL})}/\beta$
and $F_{PS} = -\ln{(Z_{PS})}/\beta$. 
For an infinitely long simulation both sets will be sampled, the time 
spent in each of them being proportional to the corresponding
partial partition function. However, since there are large
barriers between the two types of configurations, 
a simulation of
practical duration might sample just one of the sets. If the free
energies $F_{LL}$ and $F_{PS}$ are sufficiently different, one might 
expect that the simulation after an initial equilibration settles 
within the set with the lowest free energy. Hence, for a large system, a 
simulation carried out at a very low temperature would ideally show a 
discontinuous change in all calculated quantities at $V=V_{PS}$.
However, in a small system there will always be some tunneling
between $\lbrace LL \rbrace$ and $\lbrace PS \rbrace$ at any finite 
temperature if $V \approx V_{PS}$, and the transition will therefore 
appear smoothed.

The evolution of the kinetic energy and the $p$ occupation
with Monte Carlo "time" is illustrated in Fig.~\ref{time}, for a system
with $V$ larger than but close to $V_{PS}$. Each point represents an average
over a ``bin'' consisting of a large number of Monte Carlo configurations.
Initially they fluctuate around values corresponding to the Luttinger liquid
close to phase separation (see Fig.~\ref{local}), but then
change to values typical after phase separation. Smaller fluctuations 
towards the phase separated regime can also be seen, and after the 
transition there are small fluctuations back towards the Luttinger liquid 
values. Typically, low-temperature simulations rapidly 
equilibrate to either the LL or PS regime, depending on $V$, and
behavior such as that in Fig.~\ref{time} is seen
only very close to $V_{PS}$.

The energy versus $V$ is shown in Fig.~\ref{energy}, and has
a behavior typical for a level crossing. There are two approximately 
linear regimes, with different slopes. The intersection of 
lines fitted to the two sets of points indicate that $V_{PS} \approx
2.63$. The point for $V=2.675$ was actually obtained from the
part of the simulation illustrated in Fig.~\ref{time} before the 
system equilibrated into the phase separated regime. Hence, it 
represents the energy of a meta-stable Luttinger liquid.
Fig.~\ref{energy} indicates that also $V=2.65$ is on the separated side
of the transition, and hence that this simulation remained
in the meta-stable state, which is not surprising this close
to the transition.

The scenario outlined above also implies an interesting finite-temperature
behavior. For $V$ close to but smaller than  $V_{PS}$ there should be 
a regime at elevated temperatures where $F_{PS} \ll F_{LL}$, and 
the system behaves as if phase separated. Lowering the 
temperature below the lowest PS state, $F_{PS}$ increases
and the system crosses over into a Luttinger 
liquid regime where it remains down to $T=0$. This is indeed observed.
Fig.~\ref{tstruct} shows results for the charge structure 
factor and the spin susceptibility for a system with $V < V_{PS}$ 
at two temperatures. At the higher temperature 
$S_\rho (q)$ is sharply peaked at long wavelengths, which is a
clear indication of phase separation. At the
lower temperature the behavior is drastically different,
and corresponds to a uniform system. The spin susceptibility is
peaked around $q=\pi$ at the higher temperature, reflecting strong
antiferromagnetic fluctuations in the LD phase of a phase separated 
system. At the lower temperature there is a clear peak at $q=2k_F$, 
corresponding to the behavior of a Luttinger liquid with $K_\rho < 1$.
The tendency to phase separation at elevated temperatures and the subsequent
stabilization of the Luttinger liquid as the temperature is lowered
seem counterintuitive, but follow naturally from the level spectrum
we propose.

At higher doping levels the discontinuous behavior observed
for the local quantities in Fig.~\ref{local} is much less prominent,
but indications of a discontinuous phase separation transition
persist even at $\langle n\rangle = 1.50$. At $\langle n\rangle = 1.75$
the transition appears completely smooth.\cite{cuolong} 
We believe that this change
in behavior with increasing doping is due to increasing density
fluctuations in the Luttinger liquid state, which eventually, close
to $\langle n\rangle = 2$ cause the system to phase separate without
a level crossing. This has implications for the existence of dominant 
superconducting fluctuations in the neighborhood of phase separation. 
At $\langle n\rangle = 1.25$ our estimate of $K_\rho$ close to phase 
separation is $K_\rho \approx 0.8$,\cite{cuolong}
i.e. the system has dominant 
spin-density-wave fluctuations. This is also consistent with a spin 
susceptibility strongly peaked at $q=2k_F$ as shown in Fig.~\ref{tstruct}.
At $\langle n\rangle =1.50$ the fluctuations appear to become large enough 
before phase separation to result in a regime with dominant 
pairing fluctuations. This could be a finite size effect, however, 
since for $L=32$ the regime is very narrow.
For $\langle n\rangle = 1.75$ the charge fluctuations
appear to increase steadily before phase separation,
and $K_\rho > 1$ for $1.4 \alt V \alt 1.7$. In this regime we also
find indications of a gap in the spin excitation spectrum.\cite{cuolong}

It can be noted that Dagotto {\it et al.} found that the pair correlations
in small systems ($L=6$) are considerably stronger at $\langle n\rangle =1.67$
than at $\langle n\rangle =1.33$.\cite{dagotto1} This is consistent
with the scenario we have outlined above.

We conclude that the mechanism causing phase separation in the Cu-O chain
model depends on the filling. Close to half-filling it is a result of a
level crossing. There is then no associated increase in the 
long-wavelength density
fluctuations before phase separation, and therefore no dominant
pairing fluctuations. Further away from
half-filling the density fluctuations of the Luttinger liquid can
increase enough to produce pairing before phase separation. At
fillings close to two particles per unit cell these fluctuations also
appear to drive the phase separation. This intricate behavior is clearly
related to the two-band nature of the model. It would be interesting to 
explore this issue in 2D as well.

We would like to thank E. Dagotto, S. Hellberg, D. Scalapino, and 
especially C. Varma, for stimulating discussions. This work was 
supported by the Office of Naval Research under Grant 
No. ONR N00014-93-0495 (AWS) and the Research 
Council of Norway (Norges Forsk\-nings\-r{\aa}d) 
under Grants No. 110566/410 and 110569/410 (AS).

\begin{figure}
\caption{The kinetic energy $\langle \hat k\rangle$, the $p$ site
occupation $\langle n^p\rangle$, and the density of doubly occupied $p$ 
sites $\langle d^p\rangle$ vs.~$V$ for systems with $L=8-32$ unit cells 
at filling $\langle n\rangle = 1.25$.}
\label{local}
\end{figure}

\begin{figure}
\caption{Schematic outline of the level spectrum suggested to
account for the observed phase separation behavior. The  dashed 
curves represent the Luttinger liquid and its spin and charge excitations.
The solid curves represent a band of states corresponding to a phase
separated system, which crosses the Luttinger liquid ground state
at the point marked PS.}
\label{levels}
\end{figure}

\begin{figure}
\caption{The kinetic energy and the $p$ site occupation
versus Monte Carlo time. Each point represents
an average of measurements performed on a large number of configurations.}
\label{time}
\end{figure}

\begin{figure}
\caption{The energy per site vs.~$V$ for a system with 32 unit cells
at $\langle n\rangle =1.25$. Solid and open circles are for
the Luttinger liquid and phase separated regime, respectively.
The intersection of lines fitted to the points gives
$V_{PS}\approx 2.63$. Solid circles above this point indicate
metastable states.}
\label{energy}
\end{figure}

\begin{figure}
\caption{The spin susceptibility (top panel) and charge 
structure factor (bottom panel) for an $L=32$ system with 
$\langle n\rangle = 1.25$ and $V=2.6$ at two different 
inverse temperatures.}
\label{tstruct}
\end{figure}

\end{document}